\documentclass[prl,twocolumn]{revtex4}  
\usepackage{bm}
\usepackage{amssymb}
\usepackage{amsmath}
\usepackage{graphicx}

\begin{document}

\title{Gravitational Waves From Low Mass Neutron Stars}
\author{C. J. Horowitz}\email{horowit@indiana.edu} 
\affiliation{Department of Physics and Nuclear Theory Center,
             Indiana University, Bloomington, IN 47405, U.S.A}
\date{\today}
\begin{abstract}
Low mass neutron stars may be uniquely strong sources of gravitational waves (GW).  The neutron star crust can support large deformations for low mass stars. This is because of the star's weaker gravity.  We find maximum ellipticities $\epsilon$ (fractional difference in moments of inertia) that are 1000 times larger, and maximum quadrupole moments $Q_{22}$ over 100 times larger, for low mass stars than for 1.4 $M_\odot$ neutron stars.   Indeed, we calculate that the crust can support an $\epsilon$ as large as 0.005 for a minimum mass neutron star.  A 0.12 $M_\odot$ star, that is maximally strained and rotating at 100 Hz, will produce a characteristic gravitational wave strain of $h_0=2.1\times 10^{-24}$ at a distance of 1 kpc.  The GW detector Advanced LIGO should be sensitive to such objects through out the Milky Way Galaxy.  A low mass neutron star could be uniquely identified from a large observed spin down rate and its discovery would have important implications for General Relativity, supernova mechanisms, and possibly nucleosynthesis. 
\end{abstract}
\smallskip
\maketitle

Rapidly rotating neutron stars can efficiently radiate gravitational waves (GW) because they involve large masses that undergo large accelerations.  Laser interferometer GW detectors have carried out searches for such waves \cite{Searches}.  Although there have been no detections, upper limits have been set.  For example, GW radiation from the Crab pulsar is constrained to be less than 2\% of the total spin down energy budget \cite{crab, Searches}.   To generate GW one needs an asymmetry in the star.  This could be from a static ``mountain'' supported by the solid neutron star crust.  Recently we performed large scale molecular dynamics simulations of crust breaking \cite{crustbreaking}.  We find that neutron star crust is the strongest material known with a breaking stress some 10 billion times larger than steel.    This is because the long ranged Coulomb interactions and high pressure suppress many failure modes.  As a result, the crust can support large mountains, and this further motivates GW searches.

In general, a star with rotational frequency $\nu$ is expected to radiate GW with a frequency $2\nu$.  However GW radiation at frequency $\nu$ is also possible from precession or from a pinned superfluid component of the angular momentum \cite{I.Jones}.  It can be difficult to search for GW from stars with unknown $\nu$ or in unknown binary orbits.  However techniques are being developed to increase the sensitivity and reduce the large computational costs of these searches \cite{stacking}.  The sensitivity of present detectors places important constraints on the quadrupole moment $Q_{22}$, distance from earth $r$, and $\nu$ of detectable sources.  For example, for LIGO to have detected the Crab pulsar requires an ellipticity, fractional difference in moments of inertia $\epsilon=(I_{xx}-I_{yy})/I_{zz}\geq 10^{-4}$ \cite{crab}.  Ushomirsky et al. \cite{crustmonster} have calculated the maximum $Q_{22}$ (and $\epsilon$) that the neutron star crust can support.  This depends on the breaking strain $\bar\sigma$ of the crust which is equal to the breaking stress divided by the shear modulus.  Using $\bar\sigma\approx 0.1$ from our MD simulations \cite{crustbreaking}, the Ushomirsky formalism yields a maximum $\epsilon\approx 10^{-5}$ for a 1.4 $M_\odot$ neutron star, see below.   Although $10^{-5}$ is larger than previous estimates because of the large $\bar\sigma$, LIGO is not yet sensitive to these crustal mountains on the Crab.  This is, in part, because the Crab spins relatively slowly.  LIGO is already sensitive to crustal mountains on some more rapidly spinning stars. 

The core of a neutron star could involve a high density exotic solid such as a spatially nonuniform color superconducting phase \cite{exotic.solid, exotic.solid2}.  The large shear modulus of this solid could support a larger $\epsilon$ \cite{Owen, lin, haskell2007, knippel}.  However this requires that an exotic solid phase exist at neutron star densities and for this phase to become deformed.  Presumably this would have to happen shortly after the star is created in a supernova.

Alternatively, conventional neutron star crust can support larger ellipticities for low mass neutron stars.  This is because gravity is weaker on low mass stars while the strength of the crust is unchanged \cite{lowmasscalc}.  Furthermore, low mass neutron stars have very thick crusts where most of the star has a nonzero shear modulus and can help support mountains.     We expect, on very general grounds, that low mass self gravitating objects can have large deformations.  For example, low mass asteroids can be deformed, while larger mass dwarf planets are nearly spherical.   

However, it may be difficult to form low mass neutron stars directly in a supernova explosion, because low mass protoneutron stars, that are hot and lepton rich, are not gravitationally bound.  Perhaps low mass neutron stars can be formed via fragmentation during protoneutron star formation or neutron star collisions \cite{fragmentation}.  If so, low mass stars, formed via fragmentation, could be rapidly rotating and deformed.  

We speculate that rapidly rotating low mass neutron stars could generate strong gravitational waves.  We emphasize that whatever one's theoretical prejudice, the distribution of neutron star masses is an important observational question.  Gravitational wave observations could provide unique information on low mass stars.  The structure of low mass stars is interesting because their central densities are comparable to nuclear density.  As a result, their structure is closely related to properties of conventional atomic nuclei \cite{lowdensityNS}.  Low mass stars should have maximum rotation frequencies lower than the over 1 kHz frequency expected for a 1.4 $M_\odot$ star.  Low mass stars, spinning close to their maximum rate, may radiate gravitational waves with frequencies near 200 Hz.  This is where present ground based interferometers are most sensitive.  

We describe the structure of low mass neutron stars using the equation of state (EOS) of Douchin and Haensel for the inner crust and core \cite{haensel} and the original Baym, Pethick and Sutherland (BPS) EOS for the outer crust \cite{BPS}.  We discuss the sensitivity of our results to the EOS below.  The angle averaged shear modulus $\mu$ of the crust is taken from our recent molecular dynamics (MD) results \cite{shearmod},
\begin{equation}
\mu=0.1106 \frac{Z^2e^2}{a}n_i.
\label{mueq}
\end{equation}
Here the ions have charge $Z$, density $n_i$ and the ion sphere radius is $a=(3/4\pi n_i)^{1/3}$.  This result for $\mu$ is about 10\% smaller than the earlier Monte Carlo results of Ogata et al. \cite{ogata} because they neglected electron screening.

\begin{figure}[htp]
\centering
\includegraphics[width=3.6in,angle=0,clip=true] {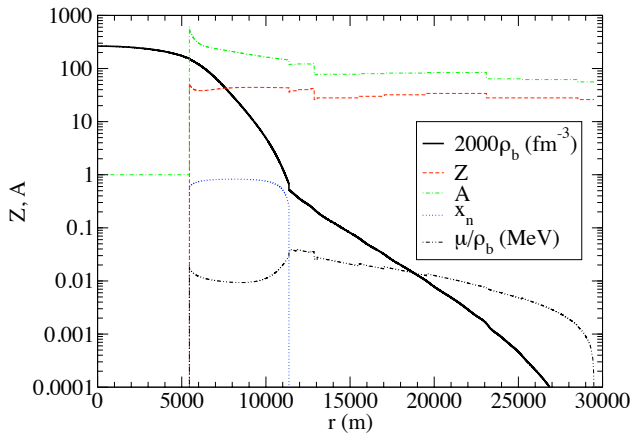}
\caption{(Color on line) Profile of a low mass 0.12 $M_\odot$ neutron star.  The charge $Z$ (red dashed line)  and mass number $A$ (green dashed-dotted line) of nuclei in the crust are plotted versus radius $r$.  Note that $A$ is arbitrarily set to one in the core.  The thick line shows the baryon density in fm$^{-3}$ multiplied by 2000 for clarity.  The dotted blue line shows the mass fraction of free neutrons $x_n$ in the inner crust.  Finally the dashed two dots line shows the ratio of shear modulus to baryon density $\mu/\rho_b$ in MeV.}
\label{Fig1}
\end{figure}

\begin{figure}[htp]
\centering
\includegraphics[width=3.6in,angle=0,clip=true] {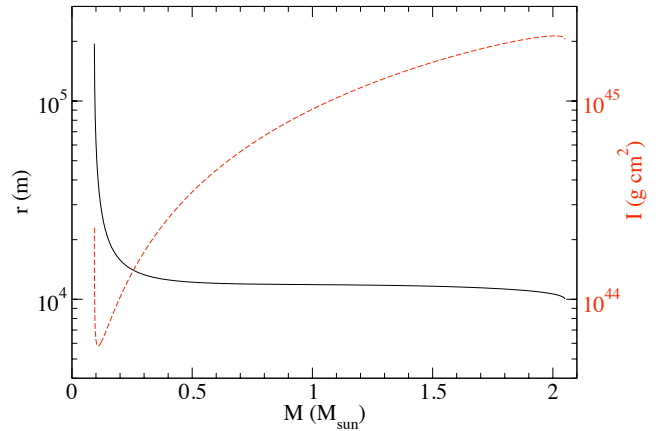}
\caption{(Color on line) Radius $r$ (solid black curve and left hand scale) and moment of inertia $I$ (dashed red curve and right hand scale) versus mass $M$ of neutron stars.}
\label{Fig3}
\end{figure}

The structure of a 0.12 $M_\odot$ neutron star is shown in Fig. \ref{Fig1}.  The charge $Z$ and mass number $A$ of nuclei in the crust are plotted.  Note that $Z$ and $A$ change continuously in the inner crust because of the statistical model empolyed for the EOS \cite{haensel}.  Also shown in Fig. \ref{Fig1} is the ratio of the shear modulus to the baryon density $\mu/\rho_b$.  The star has a radius of 29 km and is mostly solid crust with only a small liquid core.  For comparison, the structure of a conventional 1.4 $M_\odot$ star has only a thin solid crust and a large liquid core.
Figure \ref{Fig3} plots the radius of a neutron star versus its mass.  Low mass neutron stars can be much larger than 1.4 $M_\odot$ stars.


The mass quadrupole moment $Q_{22}$ of the star is,
\begin{equation}
Q_{22}=2\int d^3r \rho({\bf r}) r^2 {\rm Re}Y_{22},
\label{Q22}
\end{equation}
where $\rho({\bf r})$ is the (energy) density and $Y_{22}$ is a spherical harmonic \cite{crustmonster}.  Ushomirsky et al. have estimated the maximum $Q_{22}$ that can be supported by a crust with stress tensor $t_{ij}$\cite{crustmonster}.
In a related calculation, Haensel et al. calculated how the energy of a star rises with deformation \cite{zdunik2008}.  Knippel and Sedrakian assume the components of $t_{ij}$ are stressed just to the breaking point in a way that maximizes $Q_{22}$ \cite{knippel},
$t_{rr}=2(32\pi/15)^{1/2}\mu\bar\sigma$, 
$t_{r\perp}=2(16\pi/5)^{1/2}\mu\bar\sigma$, 
$t_{\Lambda}=6(32\pi/15)^{1/2}\mu\bar\sigma$.   Note that these relations differ by a factor of two from those in ref. \cite{crustmonster}.  We assume the breaking strain $\bar\sigma$ is,
\begin{equation}
\bar\sigma=0.1,
\label{sigmaeq}
\end{equation}
based on our large scale MD simulations \cite{crustbreaking}.  This gives a maximum $Q_{22}$ of \cite{Owen},
\begin{equation}
Q_{22}=-\bigl(\frac{32\pi}{15}\bigr)^{1/2}\int \frac{d^3r r^3 \mu\bar\sigma}{g(r)} \bigl(48-14U+U^2-\frac{dU}{d{\rm ln} r}\bigr).
\end{equation}
Here $g(r)$ is the acceleration due to gravity and $U=2+d{\rm ln}g/d{\rm ln}r$.  The ellipticity $\epsilon$ is the fractional difference in moments of inertia,
\begin{equation}
\epsilon=\frac{I_{xx}-I_{yy}}{I_{zz}}\ \ = \Bigl(\frac{8\pi}{15}\Bigr)^{1/2} \frac{Q_{22}}{I_{zz}},
\label{epsilon}
\end{equation}
where we have used Eq. \ref{Q22} for $Q_{22}$.  The moment of inertia $I=I_{zz}$, calculated in a slow rotation approximation \cite{inertia}, is shown in Fig. \ref{Fig3}.  Note that we neglect the effects of rapid rotation which can significantly distort the structure of low mass stars because of their extended crust \cite{lowmasscalc}.  The moment of inertia, of very low mass stars increases because of their large radius.

\begin{figure}[htp]
\centering
\includegraphics[width=3.6in,angle=0,clip=true] {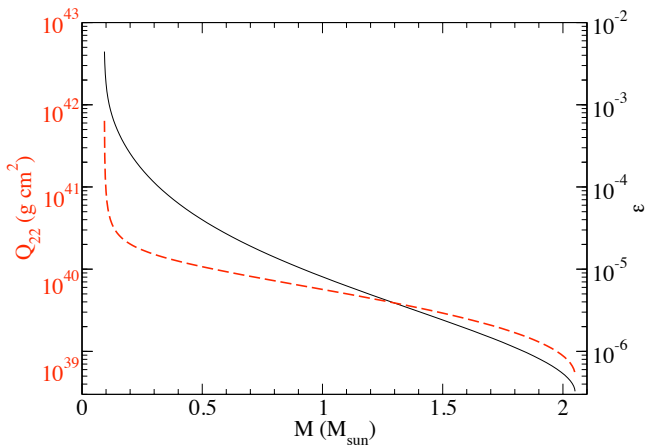}
\caption{(Color on line) Maximum ellipticity $\epsilon=\varepsilon$ (solid curve and right hand scale) and $Q_{22}$ (dashed red curve and left hand hand scale) versus mass $M$.}
\label{Fig5}
\end{figure}

The maximum $Q_{22}$ and $\epsilon$ that the crust can support are plotted in Fig. \ref{Fig5}.  The maximum $\epsilon$ for low mass stars can be as large as $5 \times 10^{-3}$.  {\it The crust can support a $Q_{22}$ that is over 100 times larger, and an $\epsilon$ that is 1000 times larger, for a low mass star than for a 1.4 $M_\odot$ star.}

The maximum rotation frequency $\nu_{\rm max}$, for a star of mass $M$ and radius $r$, has been estimated to be \cite{numax}
\begin{equation}
\nu_{\rm max} \approx 1050 \Bigl(\frac{M}{M_\odot}\Bigr)^{1/2}\Bigl(\frac{\rm 10\ km}{r}\Bigr)^{3/2}\ {\rm Hz}.
\label{numax}
\end{equation}
If this equation can be extrapolated to very low masses, it gives $\nu_{\rm max}\approx 70$ Hz for the 0.12$M_\odot$, $r=29.5$ km star in Fig. \ref{Fig1}.  Explicit two dimensional calculations of the structure of a rapidly rotating 0.13 $M_\odot$ star by Haensel et al. give 
$\nu_{\rm max}=100\ {\rm Hz}$
using the SLy EOS \cite{lowmasscalc}.  This is only slightly larger than the above estimate.  Note that this star rotating at 100 Hz is significantly oblate with the ratio of polar to equatorial radii $r_{\rm pole}/r_{\rm eq} =0.7$ \cite{lowmasscalc}.

A characteristic gravitational wave strain $h_0$, from a star with $\epsilon$ and $I$ spinning at frequency $\nu$ and a distiance $d$ away, is \cite{h0},
\begin{equation}
h_0=\Bigl(\frac{16\pi^2 G}{c^4}\Bigr)\frac{\epsilon I \nu^2}{d}.
\end{equation}
For a 0.12$M_\odot$ maximally stressed star, spinning at 100 Hz and at a distance of 1 kpc, this gives a relatively strong signal,
\begin{equation}
h_0\approx 2.1\times 10^{-24}.
\end{equation}
 The LIGO all sky search \cite{allsky} may have already ruled out such an isolated star within distances of order 1 to 3 kpc, depending on orientation.  However the sources rapid spin down rate or presence in a binary system could be complications that might greatly reduce the excluded distance.  {\it Advanced LIGO should be sensitive to these stars throughout the Milky Way Galaxy.}

Strong gravitational wave radiation will lead to a rapid spin down.  Therefore, one may be most interested in relatively young objects.  For example, Imshennik presents a speculative scenario for SN1987a that involves a very rapidly rotating core that fragments to produce a low mass neutron star \cite{Imshennik}.  We note that GW from a low mass star could be strong enough to be detected, even given the relatively large distance to SN1987a.  Therefore GW observations of SN1987a can probe the possible formation of a low mass neutron star. 


We discuss the sensitivity of our results to the EOS and approximations that we use.  First, we assume cold catalyzed matter.  In principle, the composition of accreted matter could be significantly different.  Haskell et al. have studied mountains on normal mass stars made of both accreted and non-accreted matter \cite{haskell}.  However, it seems difficult to make low mass neutron stars out of accreted matter.  Second, there could be sensitivity to the EOS of cold catalyzed matter.  Low mass neutron stars have relatively low densities, and there may be only a small range in pressure for realistic EOS at these densities.  However, the model dependence of the shear modulus, Eq. \ref{mueq} and the breaking strain, may be more important.   The crust is strongest (and can make the largest contribution to the maximum $Q_{22}$) at high densities near the base of the inner crust.  Therefore, one is sensitive to the composition $Z$, $A$ of nuclei deep in the inner crust.  This composition is set by the symmetry energy and there can be significant variations among different EOS models.  For example, Steiner and Watts find large variations in the speed of shear waves deep in the inner crust \cite{steiner}.  In addition, we expect non-spherical nuclear pasta phases at the base of the inner crust, see for example \cite{pasta}.  The impact of these phases on both the shear modulus and breaking strain should be studied further in future work.  Finally, we use the formalism of Ushomirsky et al. to estimate the maximum $Q_{22}$.  This involves the Cowling approximation and may omit some boundary terms \cite{haskell}.  Furthermore we assume a nearly spherical star, while rapidly rotating low mass stars can be significantly oblate.  These approximations should be improved in future work.  Finally, and perhaps most difficult, one should study formation mechanisms for low mass stars and ways that the star could become deformed.

We now speculate on the strongest possible sources of continuous gravitational waves for ground based detectors.  Objects significantly larger than neutron stars can not spin fast enough to radiate at frequencies where ground based detectors are most sensitive, see Eq. \ref{numax}.  Sources involving static deformations of conventional matter are constrained by the breaking stress $t\approx \mu\bar \sigma$ of neutron star crust, see Eqs. \ref{mueq},\ref{sigmaeq}.  Note that neutron star crust is the strongest material known.  Given the source's size and composition, we now optimize its mass.  Figure \ref{Fig5} clearly shows that the {\it smaller} the mass, the stronger the possible gravitational wave source.  We expect that this is a very general result.  Gravity is weaker for smaller masses.  This allows the same material strength to support larger deformations.  However there is a lower limit to neutron star masses, if they are to be bound by gravity.  The minimum neutron star mass for the EOS we use is 0.094 $M_\odot$.  We conclude that low mass neutron stars could be uniquely strong gravitational wave sources. 

There may be a useful analogy between gravitational wave searches for low mass neutron stars, and doppler shift searches for hot jupiter exoplanets.  Originally some astronomers thought that hot jupiters would be unlikely to form.  However, many planet searches are strongly biased towards the discovery of hot jupiters, because they produce large doppler shifts.  It is important to consider this bias in interpreting planet search results.  Likewise, gravitational wave searches are sensitive to deformation.  Therefore, they may be strongly biased towards the discovery of low mass neutron stars, because these stars can have large deformations.  One should consider this bias when interpreting the results of gravitational wave searches.   

To actually have a low mass gravitational wave source requires three things.  First, low mass neutron stars must be produced.  This is problematic because the minimum mass of a hot protoneutron star is much larger.  Perhaps low mass stars can be formed by fragmentation.  Second, the star must be deformed.  This deformation could have occurred when the star was formed in a violent event.  Finally, and perhaps more straight forward, the star must be rapidly rotating.  Again if the star was formed via fragmentation it is possible that the star was formed rapidly rotating.  However, the star will quickly spin down because of gravitational wave radiation.  Therefore, the system may have to be young in order for it to be still rapidly rotating.    

We strongly encourage searches for gravitational waves from low mass neutron stars.  These stars could be very strong GW sources and can be uniquely identified because of their rapid spin down rate, that is inversely proportional to the moment of inertia $I$.  Searches should be extended to include large spin down rates.  Figure \ref{Fig3} shows that $I$ can be an order of magnitude or more smaller than that for a 1.4 $M_\odot$ star.  A small observed $I$ would provide a ``smoking gun'' that the GW source is a low mass neutron star.  This star could possibly be made of exotic matter.  However the identification of low mass is unique because a larger mass object would likely collapse to a black hole before its $I$ became this small.

The GW detection of a low mass NS would have dramatic implications for General Relativity, Supernovae, and perhaps nucleosynthesis.  By proving that low mass stars are formed, presumably be fragmentation, this discovery would stress the role of rotation in at least some supernovae.  This is important because our understanding of supernova mechanisms may be significantly incomplete.  Furthermore by demonstrating the ejection of large amounts of neutron rich matter, the formation of low mass neutron stars may be related to rapid neutron capture nucleosynthesis of heavy elements. 

    
We thank Nils Andersson, Ian Jones, Ben Owen, and Lars Samuelsson for helpful discussions.   This work was supported in part by DOE grant DE-FG02-87ER40365.

\end{document}